\documentclass[aps,prl,reprint]{revtex4-2}
\usepackage{blindtext}
\usepackage[section]{placeins}

\usepackage{graphicx}
\usepackage{dcolumn}
\usepackage{bm}
\usepackage{amsmath}
\usepackage{physics}

\usepackage[mathlines]{lineno}
\modulolinenumbers[5]
\linenumbers\relax 
\begin{document}
\nolinenumbers
\preprint{AAPM/123-QED}

 \title{Measuring  Gravitational force from Femto-gram source masses}

\author{Ahmed Roman$^{1,2,3}$}
\author{Asem Hassan$^{4,5}$}
\author{Mohamed ElKabbash$^{6,*}$} 
\affiliation{$^{1}$ Department of Medical Oncology, Dana-Farber Cancer Institute, Boston, MA, USA}
\affiliation{$^{2}$Broad Institute of MIT and Harvard, Cambridge, MA, USA}
\affiliation{$^{3}$Harvard Medical School, Boston, MA, USA }
\affiliation{
$^{4}$Department of Physics, Northeastern University, Boston, MA, USA}
\affiliation{
$^{5}$Center for Theoretical Biological Physics, Northeastern University, Boston, MA, USA}
\affiliation{
$^{6}$ Research Laboratory of Electronics, MIT, Cambridge, MA, USA}
 \affiliation{$^{*}$ Corresponding author: melkabba@mit.edu}


\date{\today}

\begin{abstract}
Gravity is the weakest of all known forces. Measuring the force of gravity from micro and nano-scale source masses is an essential first step toward low-energy quantum gravity tests. In addition, measuring gravitational forces where the center-of-mass inter-distance is at the sub-mm scale extends the experimentally achievable parameter space for tests of Yukawa-like corrections to Newtonian gravity and tests for higher dimensions proposed to resolve the hierarchy problem of fundamental forces. Here, we propose an experiment using two optically trapped particles in ultrahigh vacuum conditions where the center of mass inter-distance is on the order of $10^2 nm$. In the proposed experiment, the source mass is a rotating Janus nano-particle such that the test mass (sensor) experiences a periodic gravitational potential. Using realistic experimental parameters, a signal-to-noise ratio $\geq 1$ is obtained for a Janus particle with radius $\geq 10^2 nm$ and a mass $\geq \text{10 } fg$. The proposed experiment extends the search of Yukawa corrections to gravity at $\approx 10^{-5}$ times gravity regime at $10^{2}nm $ interaction range, opens the door to low energy tests for quantum gravity, and enables experimental tests of extra-dimensional solutions to the  hierarchy problem.
\end{abstract}

\maketitle
{\em Introduction.}
Optically levitated microscopic particles trapped in ultrahigh vacuum (UHV) environments are mechanical oscillators with exquisite force sensitivity due to their weak coupling to their mechanical and thermal environment\cite{Ranjit2016}.Through optical or electronic cooling\cite{millen2020optomechanics}, the center of mass motion of a trapped nanoparticle can be cooled down from room temperature to its  ground-state of motion leading to a displacement sensitivity down to $\sim 10^{-14} \frac{m}{\sqrt{Hz}}$ \cite{magrini2021real}. This impressive displacement sensitivity can be used to measure ultra-weak forces\cite{moore2021searching}. 

Using levitated nanoparticles was proposed to measure gravitational forces at small length scales to determine corrections to Newtonian gravity \cite{Geraci2010, moore2021searching}. The source mass in the previously proposed experiments remains macroscopic in the order of a $kg$. On the other hand, measuring gravitational forces from nanoscale source masses is necessary to perform Quantum Cavendish Experiments (QCEs), i.e. experiments that test gravitational effects from quantum states of the source mass \cite{aspelmeyer2022avoid}.   The constraints on realizing QCEs requires microscopic distances between solid-state source and test masses \cite{aspelmeyer2022avoid}. To date, however, the smallest source mass measured is in the $\text{100 } mg$ range with a center-of-mass inter-distance $\approx \text{1 } mm$  which was measured through periodic modulation of the interdistance between the source and test masses \cite{Westphal2021}. On the other hand, the search for gravity-related new physics, e.g., Yukawa potential corrections \cite{dimopoulos2003probing} and higher-dimensional solutions corrections \cite{RS,AH}, requires measuring the gravitational force at small distance less than a few microns. However, decreasing  distances requires a significant decrease in the source and test mass dimensions which diminishes the signal force \cite{dimopoulos2003probing}. For example, the force of gravity scales as $R^4$ for two identical masses of radius $R$ with a negligible surface to surface distance $d\ll R$. 

\begin{figure}[!tb]
\includegraphics[width=\linewidth]{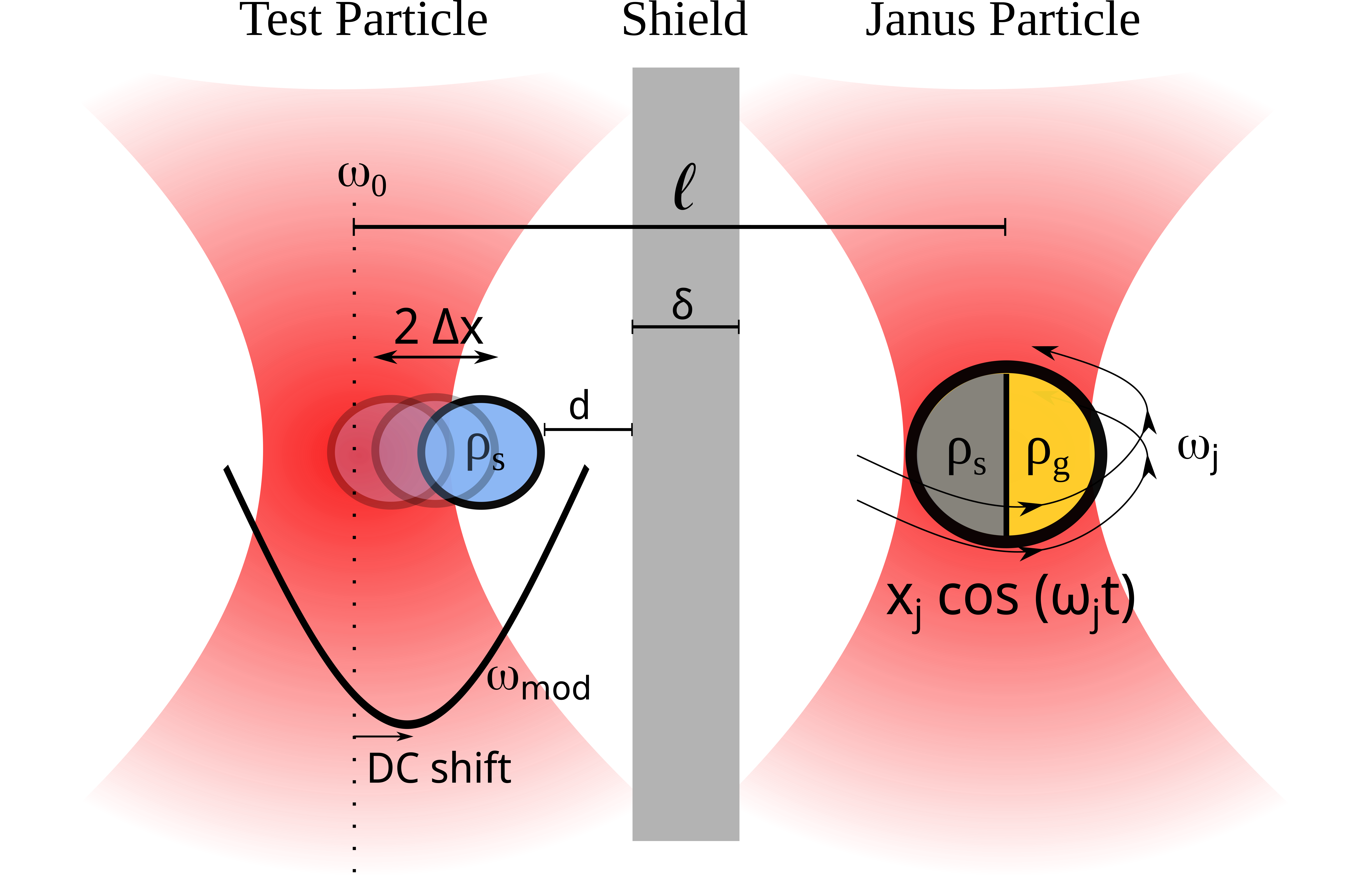}
\caption{\label{fig:epsart} Schematic of the proposed experiment. The source and test masses are optically trapped in a UHV environment. The test mass is a silica nanoparticle. The source mass is a Janus particle with one side made of gold and the other made of silica.A metallic shield with thickness $\delta$ is situated between the two masses to isolate direct Casimir force interaction and electrostatic interactions between the Janus and test particles. The Janus particle is spinning in the trap at a frequency $\omega_j$ which corresponds to a center of mass oscillation along the x axis at the same frequency. The test mass oscillates under the influence of gravity at $\omega_j$. The test mass is trapped in an optical trap potential with a natural frequency $\omega_0$. The trap's natural frequency is modified due to interacting with the shield to $\omega_{mod}$  . When $\omega_j = \omega_{mod}$, the gravity induced oscillation amplitude is amplified. }
\label{fig:model}
\end{figure}
In this work, we propose an experiment to measure the gravitational force from nano/microscale objects using experimentally feasible conditions. A signal-to-noise ratio $> 1$ for a gravitational force on the order of $ 10^{-30} N$ is realized through (i) decreasing the noise floor through cooling the test mass to its ground state of motion, (ii) using periodically modulated gravitational force through a rotating Janus particle source mass, and (iii) matching the source mass rotation frequency to the natural frequency of the test mass trapping potential. The functional form of the gravitational force at the nanoscale can be probed using the proposed experiment. Measuring gravitational forces at the nano/micro-scale opens the door for testing quantum gravity theories, as well as corrections to Newtonian gravity. We show that under experimentally achievable conditions, the proposed experiment can extend the search for Yukawa correction to gravity to the regime of $ 10^{-5}$ times gravity  for an interaction range $\lambda \geq 10^{-1}$, hence, covering most of the parameter space of theories of  Yukawa potential modified gravity and extend current bounds at 0.1 micron by up to 17 orders of magnitude. 
\cite{dimopoulos2003probing,Geraci2010}. 

{\em Experiment.} The proposed experiment is described in Figure 1. Two particles are optically trapped and levitated  under UHV conditions\cite{rudolph2022force}. The source mass is a Janus particle \cite{Schneider2013} that consists of two materials with significantly different densities, here, gold and silica. The Janus particle is spinning in the trap at a frequency $\omega_j$ \cite{reimann2018ghz}. The test mass is trapped in an optical potential with a natural frequency $\omega_0$. A metal sheet with thickness $ \delta \approx \text{50 } nm $ is present between the two masses to eliminate the Casimir and electrostatic interactions that may occur at the same frequency channel \cite{shield}. Due to the interaction between the test mass and the shield, the trap's natural frequency is modified to $\omega_{mod}$.   As a consequence of the Janus particle's oscillation, the measured displacement spectral density of the test mass will peak at the spin frequency of the source mass. When the source mass spin frequency is equal to the resonance frequency of the test mass, i.e. $\omega_j = \omega_{mod}$, the displacement amplitude is amplified and becomes detectable under experimentally achievable conditions.  It is worth noting that the shield can be made of superconducting material to eliminate the penetration of multipolar electromagnetic interactions since the field penetration depth of a superconductor is on the order of 10 nm - 100 nm \cite{szeftel2017study}. In addition, stray electromagnetic interactions can be counteracted since electric fields, unlike gravitational fields, can be neutralized \cite{priel2022dipole}. Finally, the lack of convective cooling in UHV environments can lead to  excessive heating and melting of the half-metallic Janus particle. To avoid that, the UHV chamber itself can be cooled down to low temperatures to increase the efficiency of radiative cooling (see Appendix for more details). In addition, an all-dielectric Janus particle can be used on the expense of reducing the density contrast and the displacement signal\cite{zhang2015separation, gao2020angular}. 

\begin{figure}[!tb]
\includegraphics[width=\linewidth]{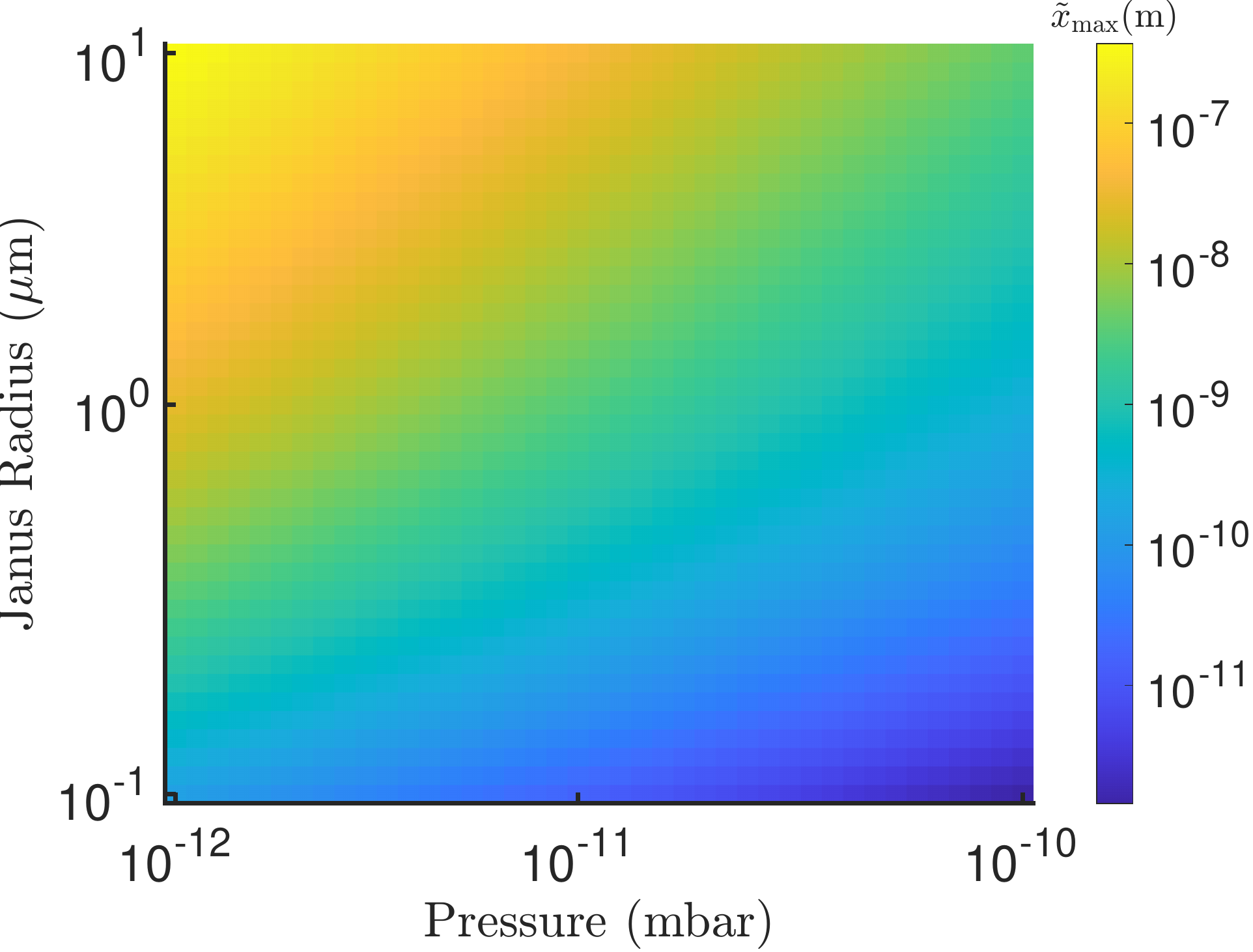}
\caption{ Maximum amplitude of oscillations $\tilde{x}_{\max}$ (in meters) reached with different chamber Pressure values (in mbar) and Janus particle radii (in meters). Parameter values used to generate this plot are: $\bar{v} =\text{430 } m/s$ at room temperature, $\rho_g = \text{19320 } kg/m^3$ and $\rho = \text{2650 } kg/m^3$. The angular frequency of the Janus particle rotation is $\omega_j = \text{10 } kHz.$ The radius of the test particle is $R= 10^{-7} \text{ }m$, and the distance between the surfaces of the two particles is $10^{-7} \text{ }m$. Thus, the distance between the centers of geometry of both particles is $\ell = R+R_j+10^{-7} \text{ }m$.}
\label{fig:maxAmpVal}
\end{figure}

{\em Model formulation.} We model the gravitational interaction of a test particle of mass $m$ with a Janus particle of mass $M_j$ separated by a distance between centers of geometry $\ell$. This distance includes a metal shield of thickness $\delta$ to isolate direct electrostatic and Casimir interactions. Placing the test particle at the origin at a distance $d$ of closest approach to the shield and distance $\ell$ to the geometric center of the Janus particle. The Janus particle is composed of a left and right hemispheres with densities $\rho_g$ and $\rho_s$. This makes the center of mass of the Janus particle displaced towards the golden hemisphere. We refer to the distance between the center of mass and the geometric center as $x_j$. As the Janus particle rotates about its $z$-axis, its center of mass  oscillates in the x direction inducing an oscillation in the position of the test particle at $x$, so that its distance from the test particle is $\ell+x_j \cos(\omega_j t)-x$. 

\begin{figure}[!tb]
\includegraphics[width=\linewidth]{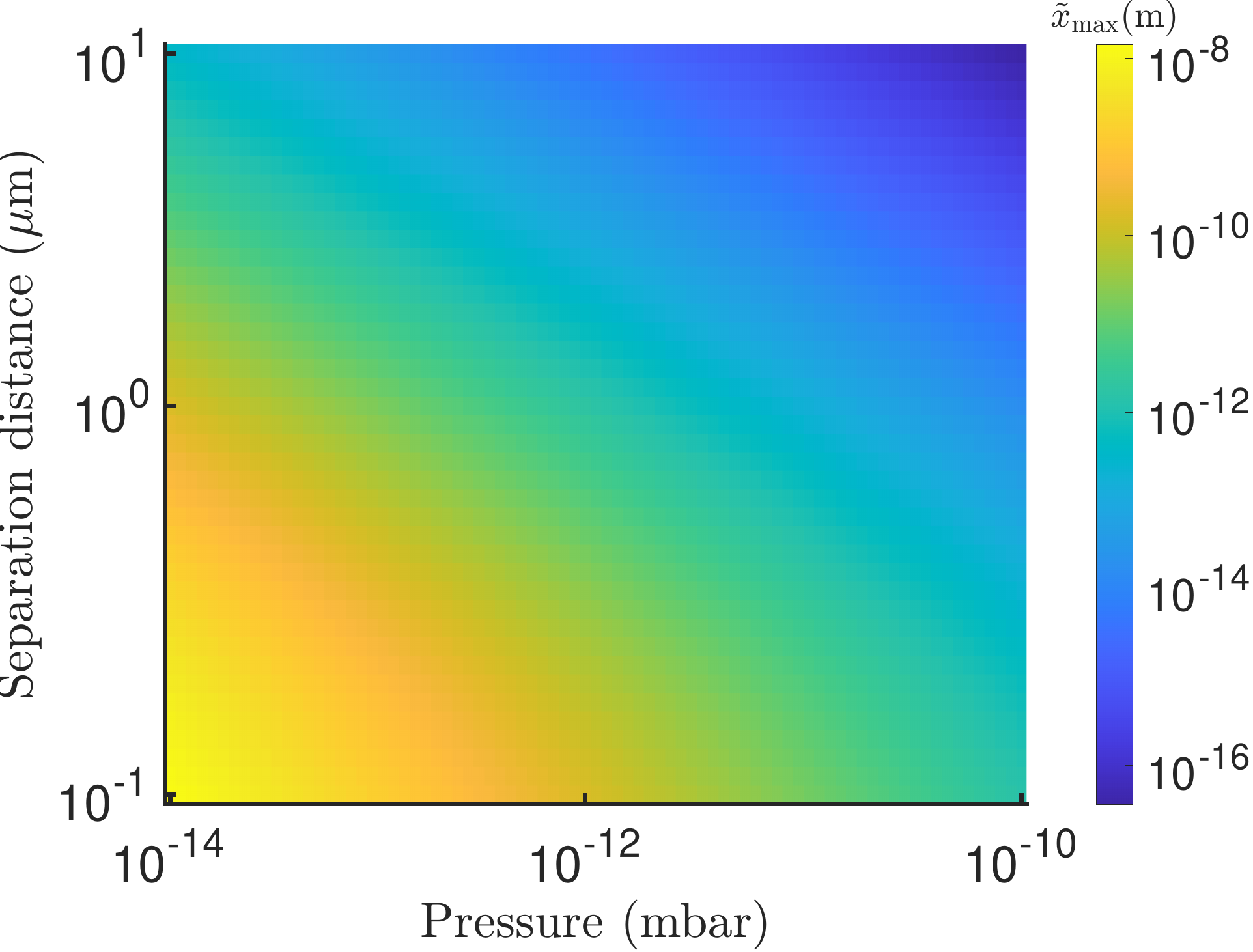}
\caption{Maximum amplitude of oscillations $\tilde{x}_{\max}$ (in meters) vs Pressure (mbar) and Separation Distance  (m). The separation distance between the particles is $\ell-R-R_j$. The Janus particle radius is $R_j= 10^{-7}m$. The parameters $\omega_j, R, \bar{v},\rho,\rho_g$ are the same as in Fig.~\ref{fig:maxAmpVal}}
\label{fig:maxAmpVal2}
\end{figure}
The test and Janus particles are placed in optical traps that act like springs with stiffness $k$ and $k_j$. We choose $k_j$ sufficiently stiff so as to effectively suppress any translational motion of the Janus particle. The test particle is subject to dipole-dipole, Casimir, and gravitational interactions $F_{dip}^s,F_{c}^s$, and $F_{g}^s$  with the shield. We refer to all these forces due to the shield as $f_s$. It is also subject to an oscillating gravitational interaction $F_{g}^j$ with the Janus particle, which induces oscillation in the position of the test particle. Letting $x$ be the position of the test particle, then the ensuing equations of motion become
\begin{equation}
x'' + \gamma x' + \omega_0^2 x = \frac{f_s + F_{g}^j}{m}+\eta_{th}(t)+\eta_{q}(t).    \label{model}
\end{equation}
where $\eta_{th}(t)$ and $\eta_{q}(t)$ are the thermal and quantum white noises respectively, and $\langle \eta_i(t)\rangle = 0$ and $\langle \eta_i(t)\eta_i(t')\rangle=\sigma_i^2 \delta(t-t')$ for $i$ thermal or quantum. For the rest of the paper we assume $\sigma_{th}\gg \sigma_{q}$, and leave the $\sigma_{th}\ll \sigma_{q}$ to the appendix. \\
{\em The small oscillations regime.} In the regime where the displacement of the test particle is small $x/d\ll1$, the dipole-dipole, Casimir interactions, and gravity forces between the test particle and metal shield are Taylor expanded in $x/d$, while the gravitational force between the test and Janus particles is Taylor expanded in $\frac{x_j\cos(\omega t)-x}{\ell}$. The constant terms in these expansions yield a DC shift in the position of the test particle, while the linear terms induce a modification of the test particle's optical trap stiffness. The ensuing equation for the DC shifted oscillator $\Tilde{x} = x-\frac{F_{DC}}{m}$ is a forced harmonic oscillator of the form
\begin{multline}
\Tilde{x}'' + \gamma \Tilde{x}' + \omega_{\text{mod}}^2 \Tilde{x} \approx  \frac{x_j F_{g}^{'j}(0)}{m}\cos(\omega_j t)+\eta(t),
\label{offsetModel}
\end{multline}
where the modified frequency $\omega_{\text{mod}}$ is given by 
\begin{equation} \omega_{\text{mod}}^2 = \omega_0^2-\frac{ F_{dip}^{'s}(0)+F_{c}^{'s}(0)
+F_{g}^{'s}(d)+F_g^{'j}(0)}{m}.
\label{approxFeq}
\end{equation}

The steady-state solution of the mean displacement of Eq.~(\ref{offsetModel}) yields the signal $\langle \Tilde{x}(t) \rangle = \Tilde{x}_{\max}\sin(\omega_j t+\phi)$
where 
\begin{equation}
\Tilde{x}_{\max} = \frac{x_jF_g^{'j}(0)}{m} \frac{1}{\sqrt{(\omega_{\text{mod}}^2-\omega_j^2)^2+ \gamma^2 \omega_j^2}}      
\label{signal}
\end{equation}
and $\phi = \tan^{-1}\left(\frac{\omega_{\text{mod}}^2-\omega_j^2}{\gamma \omega_j}\right).$ 
Note that the driving force depends only on the gravitational interaction between the Janus particle and the test particle. Consequently, the steady-state solution could be resonantly amplified to measure the weak effect of gravity.

\begin{figure}[!tb]
\includegraphics[width=\linewidth]{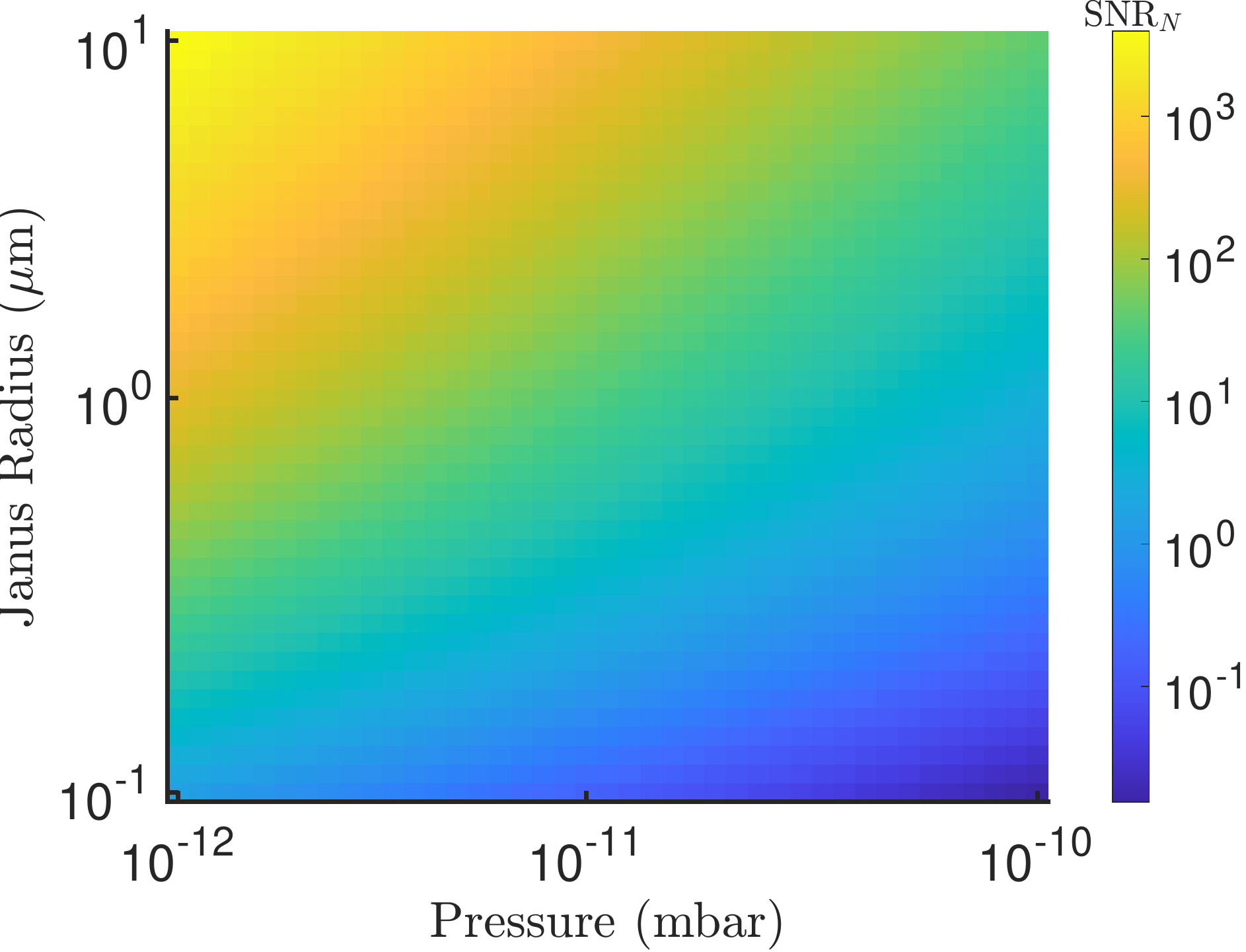}
\caption{The signal-to-noise ratio as a function of Pressure (mbar) and Janus particle radius (m). The distance between the particles is $\ell = R+R_j+10^{-7}m$. The temperature is taken from \cite{magrini2021real} as $T =7\cdot 10^{-7} K$. The parameters $\omega_j, R, \bar{v},\rho,\rho_g$ are the same as in Fig.~\ref{fig:maxAmpVal}}.
\label{fig:SNRN}
\end{figure}
The thermal noise is obtained from the equipartition theorem 
via 
$\frac{1}{2} m \omega_{\rm{mod}}^2\langle \Tilde{x}_{\rm{noise}}^2\rangle = \frac{1}{2}k_B T$
, which implies
\begin{equation}
\langle \Tilde{x}_{\rm{noise}}^2\rangle = \frac{k_B T}{m\omega_{\rm{mod}}^2}.
\label{noise}
\end{equation}
Since $\langle \Tilde{x}_{\rm{noise}} \rangle = 0$, we see that $\sigma_{th} = \sqrt{\frac{k_B T}{m\omega_{\text{mod}}^2}}.$
Using Eq.~\ref{noise} and Eq.~\ref{signal}, we define the signal to noise ratio (SNR) at $\omega_j \approx \omega_{\text{mod}}$ via
\begin{equation}
\text{SNR} = \frac{\Tilde{x}_{\max}}{\sqrt{\langle \Tilde{x}_{\rm{noise}}^2\rangle}}=\frac{x_jF_g^{'j}(0)}{\sqrt{m\gamma^2 k_B T}}.
\label{SNR}    
\end{equation}

{\em Newton's Gravity.}
If we assume that there are no corrections to Newtonian gravity at the nano-scale, then the maximum amplitude of oscillations experienced by the test particle when $\omega_{\text{mod}} \approx \omega_j$ becomes 
\begin{equation}
\Tilde{x}_{\max,N} = \frac{\pi^2 G   R R_j^4 \bar{v} \rho(\rho_g-\rho )}{32 l^3 P_{gas} \omega_j} 
\label{maxAmp}
\end{equation}
where $R$ and $R_j$ are the radii of the test and Janus particles respectively, $P_{gas}$ is the gas pressure at room temperature, $\rho_g$ and $\rho$ are densities of gold and silica respectively, and $\bar{v}$ is the rms velocity of gas particles at room temperature. In deriving Eq.~(\ref{maxAmp}), we used 
$x_j = \frac{3 (\rho_g-\rho)}{8(\rho_g+\rho)}R_j$ and the damping value $\gamma = 16P_{\text{gas}}/(\pi \bar{v} \rho R)$ found in ref. \cite{Epstein24}. This maximum displacement is detectable for a wide range of Janus particle radii and pressures $P_{\text{gas}}$ as seen in Fig.~(\ref{fig:maxAmpVal}). The Newton signal to noise ratio becomes 
\begin{equation}
\text{SNR}_{N} =  \frac{\pi^{5/2} G \bar{v}   R^{5/2} R_j^4 \rho^{3/2}(\rho_g-\rho )}{16\sqrt{3} l^3 P_{\text{gas}} k_B^{1/2} T^{1/2}}  
\end{equation}

\begin{figure}[!t]
\includegraphics[width=\linewidth]{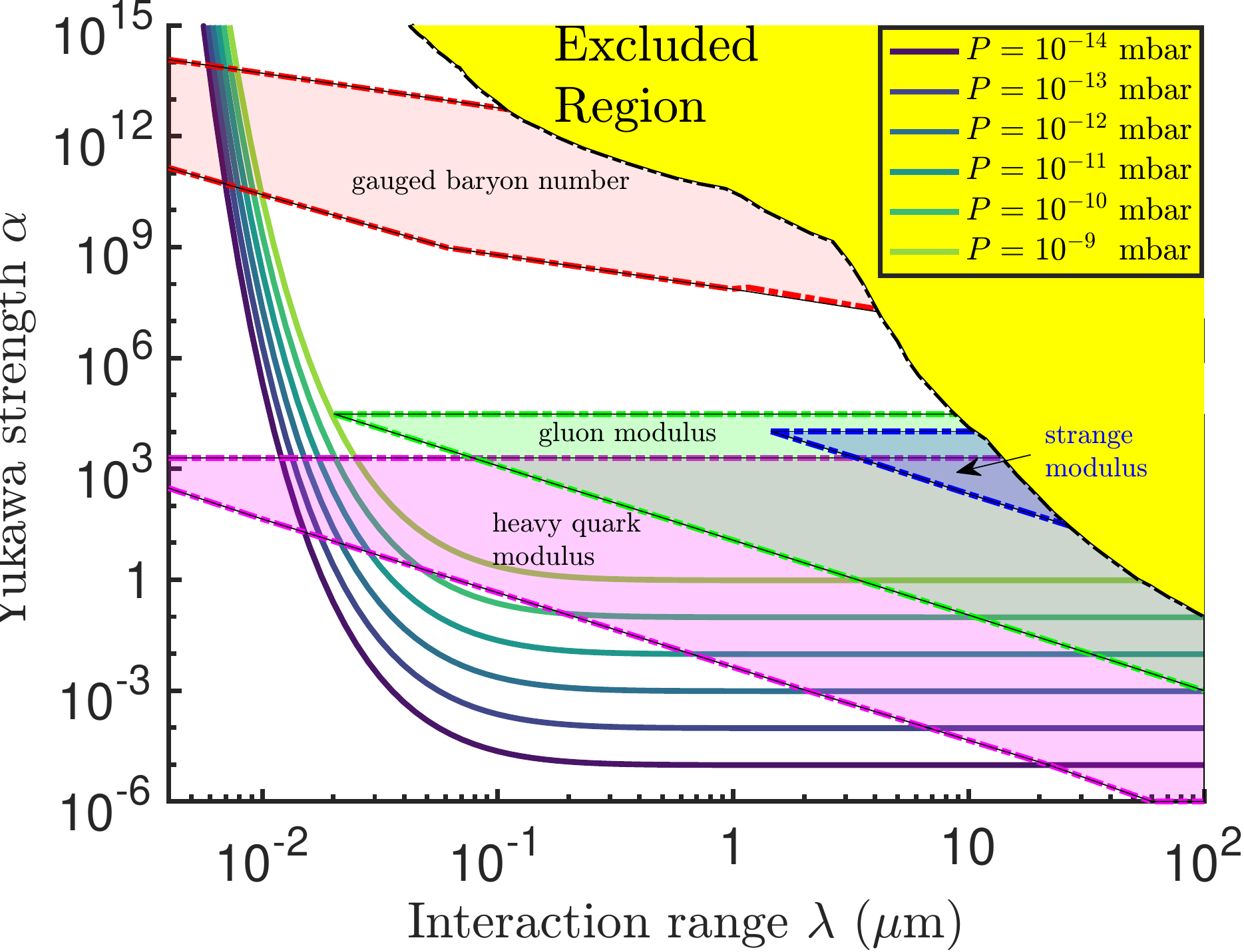}
\caption{The detectable-Yukawa boundary $(\lambda,\alpha)$ as function of Pressure (mbar). The separation distance between the particles is $\ell-R-R_j = 3\times 10^{-7} m$. The Janus particles has radius $R_j = 3\times 10^{-6}m$. The parameters $\omega_j, R, \bar{v},\rho,\rho_g$ are the same as in Fig.~\ref{fig:maxAmpVal}}.
\label{fig:YukawaBoundary}
\end{figure}
{\em Yukawa-Corrections to Newtonian Gravity}
For short-range gravitational interactions, corrections to Newtonian gravity are generally parameterized by a Yukawa-like potential of the form \cite{Dimopoulos03}
\begin{equation}
V(r) = -\frac{Gm_1m_2}{r}\left(1+\alpha e^{-r/\lambda}\right),    
\end{equation}
where the masses $m_1,m_2$ are at a distance $r$, $\lambda$ is the range of the interaction, and $\alpha$ is the relative strength the potential. If we define $Y_{\alpha,\lambda}(\ell) = \left(1+  \frac{\ell}{\lambda}+\frac{\ell^2}{2\lambda^2} \right)\alpha e^{-\ell/\lambda}.$
For this potential, the maximum amplitude of oscillation found in Eq.~(\ref{signal}) at $\omega_{\text{mod}} = \omega_j$  is given by 

\begin{align}
\Tilde{x}_{\max} &= 
\frac{\pi^2 G R R_j^4 \bar{v}\rho(\rho_g-\rho) }{32\ell^3 P_{\text{gas}} \omega_j}\left(1+Y_{\alpha,\lambda}(\ell)\right)\\
&\cong \Tilde{x}_{\max,N}+\Tilde{x}_{\max,Y}. 
\end{align}
The ensuing signal-to-noise ratio becomes 

\begin{align}
\text{SNR} &=  \frac{\pi^{5/2} G \bar{v}   R^{5/2} R_j^4 \rho^{3/2}(\rho_g-\rho )}{16\sqrt{3} l^3 P_{\text{gas}} k_B^{1/2} T^{1/2}}
\left(1+Y_{\alpha,\lambda}(\ell)\right)\\ 
&\cong \text{SNR}_N+\text{SNR}_Y.
\end{align}
Since any gravitational interaction whose strength does not lead to an $\text{SNR}_Y$ larger than one implies that Yukawa corrections are not detectable, the detectable values of $\alpha$ and $\lambda$ satisfy 
\begin{equation}
 Y_{\alpha,\lambda}(\ell)\ge 
 \frac{16\sqrt{3} l^3 P_{\text{gas}} k_B^{1/2} T^{1/2}}{\pi^{5/2} G \bar{v}   R^{5/2} R_j^4 \rho^{3/2}(\rho_g-\rho )} = \frac{1}{\text{SNR}_N}.
\end{equation}
The Yukawa-$(\alpha,\lambda)$-detection boundary satisfies

\begin{align}
\alpha &=  \frac{e^{\ell/\lambda}}{1+\frac{\ell}{\lambda}+\frac{\ell^2}{2\lambda^2}} \frac{1}{\text{SNR}_N}\approx \bigg\{ \begin{array}{ll}
\frac{1}{\text{SNR}_N}   & \quad \frac{\ell}{\lambda}\ll 1\\
\frac{1}{\text{SNR}_N}\frac{2\lambda^2}{\ell^2}e^{\frac{\ell}{\lambda}}&\quad \frac{\ell}{\lambda}\gg 1. 
\end{array}
\end{align}
Any theoretical massive-force carrying particles (moduli) in the regions above the Yukawa-detection boundaries shown in Fig.~\ref{fig:YukawaBoundary} could be probed using the current proposal. In particular, for pressures near $10^{-14}$ mbar, we can detect forces that are $\mathcal{O}(10^{-5})$ smaller than the force of gravity at an interaction range of $10^2 nm$, which presents an improvement of $\mathcal{O}(10^{18})$ on the experimentally excluded Yukawa-strength $\alpha$.    

{\em Direct tests of extra-dimensions and the Hierarchy problem}
To resolve the mass hierarchy problem of the standard model (SM) between the weak scale and gravity, proposals of quantum gravity theories lower the scale of quantum gravity from the Planck scale $\sim 10^{16} \text{TeV}$ to about $1 \text{TeV}$. Models with large extra dimensions, e.g. the Randall-Sundrum (RS) and the Arkani-Hamed–Dimopoulous–Dvali (ADD) models \cite{RS,AH}, induces modification of the gravitational force. The 5D RS model predicts gravitational corrections of the form $V_{\text{RS}}(r) = \frac{GmM}{r}\left(1+\frac{1}{r^2 k^2}\right)$, which yields a maximum mean displacement of $\Tilde{x}_{\max} = \Tilde{x}_{\max,N}(1+6/k^2\ell^2)$
and $\text{SNR} = \text{SNR}_N (1+6/k^2\ell^2)$. The maximum detectable $k$ becomes 
\begin{equation}
k_{\max}^2 =  \text{SNR}_N (6/\ell^2).
\end{equation}
At $10^{-12}$ mbar pressure and $R_j = 800$ nm, $R = 100$ nm, and all other parameters as in fig.~\ref{fig:SNRN}, we can probe distances $k^{-1}\approx 2.5\times 10^{-8}m$. While the LHC can probe much smaller distances of $10^{-19}m$ or energies of order $\text{13 } TeV$, its tests of extra-dimensional gravity remain indirect. For example, the LHC searches for increased productions of top-quarks or Kaluza-Klein excitation of the graviton predicted by the RS model \cite{Olaf2019,CMS2007,CMS2016}. Direct tests of gravity at the sub-micron scale can detect extra-dimensions even if the underlying theory does not predict particles at LHC accessible energy scales.         


{\em Conclusions.} Levitated nano-particles cooled to their quantum ground state of motion represent one of the largest non-classical systems ever realized. Introducing rotating Janus particles is a new addition to the experimental toolkit of levitation optomechanics as they provide a mechanical system that oscillates at frequencies $\leq GHz$. We showed that the gravitational force of a Janus particle source mass could be measured by optically levitated nanoparticles cooled and trapped in ultrahigh vacuum conditions. In addition to the proposed experiment, rotating Janus particles can be used to probe Casimir forces and even to demonstrate the Dynamic Casimir Effect. Janus particles are programmable mechanical oscillators making them attractive for establishing coherent superpositions of massive objects.

\begin{acknowledgments}

\end{acknowledgments}

\bibliography{MeasuringGravity}

\section*{Appendix}
{\em The perturbative effect of Gravity on the test particle.}
The gravitational force between the test and Janus particles is $\vec{F}_g = \frac{G m_j m}{|\overrightarrow{r_j-r}|^3}\overrightarrow{r_j-r}$ where $\vec{r}_j = (\ell+x_c \cos(\omega_j t),x_c \sin(\omega_j t))$ and $\vec{r}=(x,y)$ are the position vectors of the Janus and test particles respectively. Taylor expanding the force to first order in $\frac{x_c \cos(\omega_j t)-x}{\ell}$ and $\frac{x_c \sin(\omega_j t)-y}{\ell}$ yields two decoupled equations of motion in the $\hat{x}$ and $\hat{y}$ directions representing two driven harmonic oscillators. While solutions to the equations of motion up to first order are accurate up to $\mathcal{O}(x_c/l)$, $\mathcal{O}(d/l)$, and $\mathcal{O}(x/l)$, numerical solutions to these equations can obtain solutions whose accuracy is noise limited.
\\
{\em Calculating the thermal average of $\Tilde{x}^2$.}
For the one-dimensional harmonic oscillator whose Hamiltonian is 
\begin{equation}
H = \frac{p^2}{2m}+\frac{m\omega^2x^2}{2}+f x \cos(\omega t).
\end{equation} The average over one cycle of H yields $\bar{H}$ which is the classical Hamiltonian $\bar{H} = \frac{p^2}{2m}+\frac{m\omega^2x^2}{2}$. The thermal average of $\Tilde{x}^2$ for $\bar{H}$
\begin{equation}
\langle \Tilde{x}^2\rangle = \frac{\int \Tilde{x}^2 e^{-\frac{\beta m\omega^2\Tilde{x}^2}{2} }d\Tilde{x} \int e^{-\frac{\beta p^2}{2m}}dp}{ \int e^{-\frac{\beta m\omega^2\Tilde{x}^2}{2} }d\Tilde{x} \int e^{-\frac{\beta p^2}{2m}} dp}= \frac{k_B T}{m \omega^2}.
\end{equation}

{\em Deriving power spectral density of thermal noise.}
Letting $\chi(\omega)$ be the Fourier transform of $\Tilde{x}(t)$ and taking the Fourier Transform of Eq.~\ref{offsetModel} we find that
\begin{equation}
\chi(\omega) =\frac{1}{m} \frac{\hat{F}(\omega)+\hat{\eta}(\omega)}{(\omega_{\rm{mod}}^2-\omega^2)+i\gamma \omega}
\end{equation}
where $\hat{F}(\omega)$ and $\hat{\eta}(\omega)$ are the Fourier transform of $F(t) = x_j F_g'(0) \cos(\omega_j t)$ and $\eta(t)$ respectively. From this, we can obtain the power spectrum of the noise component 
\begin{equation}
S_{\Tilde{x}_{\rm{noise}}\Tilde{x}_{\rm{noise}}}=S_{\eta \eta}(\omega)|\chi(\omega)|^2  = 
\frac{S_{\eta \eta}}{m^2[(\omega_{\rm{mod}}^2-\omega^2)^2+\gamma^2 \omega^2]}.
\end{equation}
 Since $\eta(t)$ is white noise, $S_{\eta \eta}$ is a constant independent of the frequency $\omega$. From the power spectrum, we calculate the variance of the noise in the position (since $\langle \Tilde{x}_{\rm{noise}}\rangle=0$) as
 
 \begin{equation}
\langle \Tilde{x}_{\rm{noise}}^2\rangle =\frac{S_{\eta \eta}}{2 \pi m^2} 
\int_{-\infty}^{\infty} 
\frac{d\omega}{(\omega_{\rm{mod}}^2-\omega^2)^2+\gamma^2 \omega^2}.
\label{noiseIntegral}
 \end{equation}

{\em Normalizing the power spectrum} 
To determine the value of the integral in Eq.~\ref{noiseIntegral}, we close the contour in the upper half-plane with a half-circle at infinity. We find that there are two poles in the upper half-plane at 
\begin{equation}
\omega_{\pm} = \frac{i\gamma \pm \sqrt{4\omega_{\rm{mod}}^2-\gamma^2}}{2}.
\end{equation}
The value of the integral in Eq.~\ref{noiseIntegral} is the same as the contour integral because the integral over the arc at infinity ($|\omega|\rightarrow \infty$) renders the integrand zero. The contour integral is determined by the residues of the power spectrum of $\Tilde{x}_{\rm{noise}}$ though 
\begin{align}
\langle \Tilde{x}_{\rm{noise}}^2\rangle &= \frac{1}{2\pi}\oint\limits_{C}S_{\eta \eta}(\omega)|\chi(\omega)|^2 \,\mathrm{d}\omega \nonumber \\
&= i\left( \Res_{\omega = \omega_+}S_{\Tilde{x}_{\rm{noise}}\Tilde{x}_{\rm{noise}}}+\Res_{\omega = \omega_-}S_{\Tilde{x}_{\rm{noise}}\Tilde{x}_{\rm{noise}}}\right) \nonumber \\
&= \frac{S_{\eta \eta}}{2m^2 \omega_{\rm{mod}}^2\gamma}
\label{variance}
\end{align}
where the residues are given by 
\begin{equation*}
\Res_{\omega = \omega_{\pm}}S_{\Tilde{x}_{\rm{noise}}\Tilde{x}_{\rm{noise}}} = 
\frac{\pm S_{\eta \eta}}{ m^2 \gamma (-\gamma \sqrt{4\omega_{\rm{mod}}^2-\gamma^2}\pm i(4\omega_{\rm{mod}}^2-\gamma^2))}.     
\end{equation*}
From Eq.~\ref{noise} and Eq.~\ref{variance}, we find that 
$S_{\eta \eta} = 2m\gamma k_B T.$ The resulting power spectrum becomes 
\begin{equation}
S_{\Tilde{x}_{\rm{noise}}\Tilde{x}_{\rm{noise}}}= 
\frac{2\gamma k_B T}{m[(\omega_{\rm{mod}}^2-\omega^2)^2+\gamma^2 \omega^2]}.
\end{equation}

{\em Quantum and classical noises.}
The eigenenergies of the quantum harmonic oscillator are given by $E = \hbar \omega(n + \frac{1}{2})$. In the ground state, energy $\frac{\hbar \omega}{2}$ is shared between the expected kinetic and potential energies i.e. $\langle V\rangle = \langle K\rangle = \frac{\hbar \omega}{4}$. The equipartition theorem implies $\frac{1}{2} m\omega^2 \langle x_{\text{noise}}^2\rangle_{\text{QM}} = \frac{\hbar \omega}{4}$ i.e. 
\begin{equation}
\sqrt{\langle x_{\text{noise}}^2\rangle_{\text{QM}}} = \sqrt{\frac{\hbar}{2m\omega}}.
\end{equation}
Comparing this to the thermal noise ${\langle x_{\text{noise}}^2\rangle_{\text{Th}}} = \sqrt{\frac{k_B T}{m\omega^2}}$, we see that 
thermal noise is bigger than quantum noise in the regime $\frac{\hbar \omega}{2} \ll k_B T$. When the two are of the same order of magnitude, the noises add and the SNR diminishes by a factor of $\sqrt{2}$. In the opposite limit $\frac{\hbar \omega}{2} \gg k_B T$, quantum noise dominates and the SNR is    
\begin{equation}
\text{SNR} = \frac{\Tilde{x}_{\max}}{\sqrt{\langle \Tilde{x}_{\rm{noise}}^2\rangle_\text{QM}}}.
\end{equation}

{\em Radiative cooling of trapped nanoparticles.} Metallic nanoparticles are routinely trapped under ambient conditions \cite{lehmuskero2015laser}. Heating of the trapped nanoparticles from the trapping laser is commonly observed. However, convective cooling is sufficient to cool the particle down. We assume the trapping laser equally heats a trapped particle under ambient or UHV conditions. The convective heat flux is given by 
\begin{equation}
Q_{\text{convective}}= h (T_{\text{particle}}-T_{\text{medium}})
\end{equation}
where the heat transfer coefficient $h \approx 10 \ W m^{-2} K^{-1}$ \cite{elkabbash2022radiative, chen2016radiative}. Under ambient conditions and assuming the trapped particle temperature increases by $50\ K$ as discussed in \cite{lehmuskero2015laser}, then $Q_{\text{convective}} \approx 500\ W m^{-2}$. On the other hand, the total radiated heat flux is given by Stefan-Boltzmann law,
\begin{equation}
Q_{\text{radiative}}= \sigma (T_{\text{particle}}^4-T_{\text{medium}}^4)
\end{equation} 
where $\sigma = 5.62 \times 10^{-8} W m^{-2} K^{-4}$ is Stefan-Boltzmann coefficient. If we cool the vacuum chamber to liquid Nitrogen temperature ($\approx 70 K$), then  $Q_{\text{radiative}} \approx 700 \ W m^{-2} $. Consequently, radiative cooling can efficiently cool down trapped metallic nanoparticles.

\end{document}